\documentclass[aps,prd,preprint,nofootinbib,superscriptaddress]{revtex4}
\usepackage{color}
\usepackage{graphicx}
\usepackage{psfrag}

\newcommand{\bp}{\mbox{\boldmath $p$}}
\newcommand{\bk}{\mbox{\boldmath $k$}}
\newcommand{\br}{\mbox{\boldmath $r$}}
\newcommand{\bx}{\mbox{\boldmath $x$}}
\newcommand{\by}{\mbox{\boldmath $y$}}
\newcommand{\bz}{\mbox{\boldmath $z$}}

\begin{document}
\preprint{UTHEP-650}
\preprint{INT-PUB-12-059}
\preprint{RIKEN-QHP-55}

\title{Construction of energy-independent potentials  above inelastic thresholds in quantum field theories}

\newcommand{\TsukubaA}{
  Graduate School of Pure and Applied Sciences, University of Tsukuba,
  Tsukuba 305-8571, Japan
}

\newcommand{\TsukubaB}{
Center for Computational Sciences, University of Tsukuba, Tsukuba 305-8577, Japan
}

\newcommand{\TsukubaC}{
Kobe Branch, Center for Computational Sciences, University of Tsukuba, in RIKEN Advanced Institute for Computational Science(AICS), PortIsland, Kobe 650-0047, Japan
}

\newcommand{\Tokyo}{
Department of Physics, The University of Tokyo, Tokyo 113-0033, Japan
}

\newcommand{\Riken}{
Theoretical Research Division, Nishina Center, RIKEN, Wako 351-0198, Japan
}

\newcommand{\Nihon}{
Nihon University, College of Bioresource Sciences, Kanagawa 252-0880, Japan
}

\author{Sinya~Aoki}
\affiliation{\TsukubaA}
\affiliation{\TsukubaB}

\author{Bruno~Charron}
\affiliation{\Tokyo}

\author{Takumi~Doi}
\affiliation{\Riken}

\author{Tetsuo Hatsuda}
\affiliation{\Riken}

\author{Takashi~Inoue}
\affiliation{\Nihon}

\author{Noriyoshi~Ishii }
\affiliation{\TsukubaC}

\begin{center}
\includegraphics[width=0.35\textwidth]{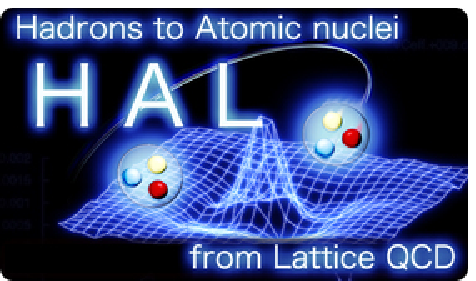}
\end{center}

\begin{abstract}
We construct energy independent but non-local potentials above  inelastic thresholds,
in terms of Nambu-Bethe-Salpeter wave functions
defined in quantum field theories such as QCD. As an explicit example, we consider $NN\rightarrow NN + n\pi$ scattering processes for $n=0,1,2,\cdots$. We show an existence of energy-independent coupled channel potentials with a non-relativistic approximation, where momenta of all particles are small compared with their own masses.
In the case of two-body inelastic scatterings such as $\Lambda\Lambda\rightarrow \Lambda\Lambda, N\Xi,\Sigma\Sigma$, on the other hand, we show that energy-independent potentials can be constructed without relying on non-relativistic approximations.
We also propose a method to extract these potentials using time-dependence of general correlation functions. 
\end{abstract}

\maketitle

\section{Introduction}
\label{sec:introduction}
It is important to understand hadronic interactions such as nuclear forces from the point of view of their constituents, quarks and gluons, whose dynamics is described by Quantum Chromodynamics (QCD).
Since the running coupling constant in QCD becomes large at hadronic scale, however,
non-perturbative methods such as the lattice QCD combined with numerical simulations must be employed to
investigate this problem systematically. Conventionally the finite size method\cite{Luscher:1990ux} has been employed to extract the scattering phase shift, but the method is applicable only below the inelastic (particle production) threshold. See Refs.\cite{Hansen:2012bj, Briceno:2012yi} for an extension of this method to multi-channel systems.

Recently an alternative method was proposed to investigate hadronic interactions and it has been successfully employed to extract the potential between nucleons  below inelastic thresholds\cite{Ishii:2006ec, Aoki:2008hh, Aoki:2009ji}.
Since then, this method has been applied to
other more general hadronic interactions such as baryon-baryon interactions\cite{Nemura:2008sp, Nemura:2009kc,Inoue:2010hs,Inoue:2010es,Inoue:2011ai}, meson-baryon interactions\cite{Ikeda:2010sg,Kawanai:2010ev} and three nucleon forces\cite{Doi:2010yh,Doi:2011gq}.  See Refs.~\cite{Aoki:2011ep,Aoki:2012tk} for reviews of recent activities.

In the method, called the HALQCD method, a potential between composite particles is defined in quantum field theories such as QCD.
There are two important properties to be proven  in quantum field theories, in order to define the potential, which is a quantum mechanical object.
Let us explain the HALQCD method and these two important properties, by considering the $NN$ potential as an example.
We first introduce the equal-time Nambu-Bethe-Salpeter (NBS) wave function\cite{Balog:2001wv} in the center of mass system defined by
\begin{eqnarray}
\varphi_{W,c_0}(\bx) &=& \langle 0 \vert T\left\{N(\br, 0) N(\br + \bx, 0) ]\right\}\vert NN,W, c_0\rangle_{\rm in}
\end{eqnarray}
where $\langle 0 \vert={}_{\rm out}\langle 0 \vert={}_{\rm in}\langle 0 \vert$ is the QCD vacuum (bra-)state,
$ \vert NN, W, c_0\rangle_{\rm in} $ is the two-nucleon asymptotic in-state at the total energy $W=2\sqrt{\bk^2+m_N^2}$ with the nucleon mass $m_N$ and the relative momentum $\bk$, $c_0$ represent quantum  numbers other than $W$ such as helicity of nucleons and the direction of $\bk$,  $T$ represents the time-ordered product, and $N(x)$ with $x=(\bx,t)$ is the nucleon operator defined by $ N(x) =\varepsilon_{abc} (u_a(x)^T C\gamma_5 d_b(x) ) q_c(x)$
with the charge conjugation matrix $C$ and $q(x) =(u(x),d(x))^T$.  Note that a different choice for $N(x)$ is possible as long as $N(x)$ can annihilate the 1-particle nucleon state and the difference leads to a difference in NBS wave functions defined from them. 
Note also that $N(x)$ and $\varphi(x)$ implicitly have spinor and flavor indices. 

An important property of the NBS wave function for the definition of the potential is that,
as the distance between two nucleon operator, $x= \vert \bx \vert$, becomes large, the NBS wave function satisfies the free Schr\"odinger( or equivalently the free Klein-Gordon)  equation, 
\begin{eqnarray}
\left( E_W - H_0 \right) \varphi_{W, c_0}(\bx) &\simeq& 0, \qquad  E_W=\frac{\bk^2}{2\mu},\quad H_0= \frac{-\nabla^2}{2\mu}
\label{eq:asymptotic}
\end{eqnarray}
where $\mu = m_N/2$ is the reduced mass. In addition, the asymptotic behavior of the NBS wave function is described in terms of the phase $\delta$ determined by the unitarity of the $S$-matrix,  $S=e^{2i\delta}$,
in QCD  (or the corresponding quantum field theory).  This has been shown originally for the elastic $\pi\pi$ scattering
\cite{Lin:2001ek, Aoki:2005uf}, where the partial wave of NBS wave function for the orbital angular momentum $L$ becomes
\begin{eqnarray}
\varphi_W^L &\simeq& A_L \frac{\sin ( k x - L\pi/2+\delta_L(W) )}{k x}
\label{eq:phase}
\end{eqnarray}
as $x\rightarrow \infty$ at $W< W_{\rm th} = 4m_\pi$ (the lowest inelastic threshold).
The asymptotic behavior of the NBS wave function for the elastic $NN$ scattering has been derived in Ref.~\cite{Ishizuka2009a}.  
The asymptotic behavior of the NBS wave function such as eq.~(\ref{eq:phase}) is the first important property that motivates the definition of the potential in QCD.

The (non-local) potential between two nucleons below the inelastic threshold is defined by the  equation that
\begin{eqnarray}
\left( E_W - H_0 \right) \varphi_{W, c_0}(\bx) &=& \int d^3y\, U(\bx,\by)\ \varphi_{W, c_0}(\by) 
\label{eq:defU}
\end{eqnarray}
at $W< W_{\rm th} = 2 m_N + m_\pi$.  
In general, the non-local potential $U(\bx,\by)$ could depend on the energy $W$\cite{Luscher:1990ux}.
As we will show, however, an energy-independent potential $U(\bx,\by)$ such that eq.~(\ref{eq:defU}) is satisfied for all $W< W_{\rm th}$ can be constructed. 
Therefore, if we solve the Schr\"odinger equation with this potential in the infinite volume, its solutions automatically provide  correct phase shifts in QCD at all $W < W_{\rm th}$ by construction.
An existence of the $W$-independent potential $U$ is the second important property to define the potential 
in the HAL-QCD method.

Using the inner product $(f,g) =\int d^3 x\, \overline{f (\bx)} g(\bx) $ where $\overline{f}$ is the complex conjugate of $f$,  we introduce a norm kernel defined by ${\cal N}_{W_1c_0,W_2d_0} = ( \varphi_{W_1,c_0}, \varphi_{W_2,d_0})$. Since NBS wave functions at $W<W_{\rm th}$ are in general linearly independent\footnote{
This holds at least for $ (W_1,c_0)\not= (W_2,d_0)$ in the sufficiently large volume.
Even if some wave functions accidentally become linearly-dependent in small volume,
we can remove them, so that our construction of the energy-independent potential remains the same. }, 
an inverse ${\cal N}^{-1}$ exists and it satisfies for $W_1,W_2 < W_{\rm th}$ \footnote{We first consider the finite volume, so that $W_1,W_2$ take discrete values. We then take the infinite volume limit, so that $\delta_{W_1,W_2}$ and $\sum_W$ should be replaced by $\delta(W_1-W_2)$ and $\int dW$, respectively.} 
\begin{eqnarray}
\sum_{W < W_{\rm th},c_0} {\cal N}^{-1}_{W_1d_0,Wc_0}\ {\cal N}_{Wc_0, W_2e_0} &=&
\sum_{W< W_{\rm th},c_0} {\cal N}_{W_1d_0,Wc_0}\ {\cal N}^{-1}_{Wc_0, W_2e_0} =\delta_{W_1,W_2}\delta_{d_0,e_0}.
\end{eqnarray} 
Using the inverse norm kernel,  we define a ket vector $\vert \varphi_{W,c_0}\rangle$ as
$\langle \bx \vert \varphi_{W,c_0}\rangle\equiv\varphi_{W,c_0}(\bx)$ and its conjugate bra vector $\langle \psi_{W,c_0}\vert$ as
\begin{eqnarray}
\langle \psi_{W,c_0}\vert \bx \rangle &\equiv & \sum_{W_1<W_{\rm th},d_0} {\cal N}^{-1}_{Wc_0,W_1d_0}
\overline{\varphi_{W_1,d_0}(\bx)},
\quad
\langle \psi_{W_1,c_0}\vert \varphi_{W_2,d_0}\rangle = \delta_{W_1,W_2}\delta_{c_0,d_0} ,
\end{eqnarray}
so that
the non-local potential can be constructed as\cite{Aoki:2009ji} 
\begin{eqnarray}
U &=& \sum_{W< W_{\rm th},c_0} \left(E_{W}- H_0\right) \vert \varphi_{W,c_0} \rangle 
\langle\psi_{W,c_0} \vert,
\label{eq:potentialW}
\end{eqnarray}
since it is easy to see that it satisfies the Scr\"odinger equation (\ref{eq:defU}) as
\begin{eqnarray}
 U \vert \varphi_{W,c_0}\rangle &=& \sum_{W_1< W_{\rm th},d_0} \left(E_{W_1}- H_0\right)\vert \varphi_{W_1,d_0} \rangle \langle \psi_{W_1,d_0}\vert \varphi_{W,c_0}\rangle  
= \left(E_W- H_0\right) \vert\varphi_{W,c_0} \rangle
\end{eqnarray}
as long as $W < W_{\rm th}$. 
It should be noted that the non-local potential which satisfies eq.~(\ref{eq:defU}) at $W< W_{\rm th}$ is not unique. For example, we may add an arbitrary term proportional to  $(1- P) $  to the non-local potential $U$ without affecting eq.~(\ref{eq:defU}), where the projection is defined by
$P =\displaystyle\sum_{W<W_{\rm th},c_0} \vert \varphi_{W.c_0}\rangle \langle \psi_{W,c_0}\vert$.

The purpose of this paper is to construct an energy independent (non-local) potential which satisfies an appropriate Schr\"odinger equation at low energy but above inelastic thresholds in quantum field theories.
To make our argument more concrete, we mainly consider the $NN$ scattering in this paper. 

In Sec.~\ref{sec:construction}, we demonstrate that energy-independent potentials can be constructed above inelastic thresholds if the total energy is small enough such that the non-relativistic approximation is applicable.
In Sec.~\ref{sec:NNpi} we  consider $NN\rightarrow NN, NN\pi$ scattering as a simplest case, where the total energy $W$ is above $2m_N + m_\pi$ but below $2m_N +  2m_\pi$. 
In Sec.~\ref{sec:general},
we  generalize our construction to larger value of $W$ where the $NN\rightarrow NN + n \pi$ scattering for higher integer $n$ can occur. In this case, momenta of all particles must be still non-relativistic. 
In Sec.~\ref{sec:special}  we treat a special case of inelastic scattering such as $AB \rightarrow AB, CD$, where
non-relativistic approximation is not required to construct energy independent coupled channel potentials. 
In Sec.~\ref{sec:t-dep}, using results obtained in the previous section, we generalize the time dependent method for the extraction of the potential\cite{HALQCD:2012aa} to the case at $W \ge W_{\rm th}$, in order to treat  inelastic processes.
Conclusions and discussions  are given in Sec.~\ref{sec:conclusion}.
In Appendix~\ref{app:comp}, we compare the construction of the energy-independent potential above inelastic threshold given in the main text  with other possible variations.   

\section{Construction of energy-independent potentials above inelastic thresholds}
\label{sec:construction}
We here construct energy-independent (non-local) potentials even above inelastic thresholds for the $NN$ scattering in the center of mass system. 
In this report we only consider pion productions whose $n$-th threshold energy is given by $W_{\rm th}^n = 2m_N + n\times m_\pi$ with $m_\pi$ being the pion mass. Extensions to other particle productions such as $N\bar N$ or $K\bar K$,  etc. are straightforward. 

We introduce energy intervals defined by $\Delta_n = [ W_{\rm th}^n, W_{\rm th}^{n+1})$ for $n=0,1,2,\cdots$. Given the total energy $W$, the kinetic energy of the $NN + n\pi$ system is denoted by $E^n_{W}$, which is given by
\begin{eqnarray}
E^n_{W} &=& \frac{\bp_1^2}{2m_N} +\frac{\bp_2^2}{2m_N} +\sum_{i=1}^n \frac{\bk_i^2}{2m_\pi}, \quad
W= \sqrt{m_N^2+\bp_1^2} +\sqrt{m_N^2+\bp_2^2}+\sum_{i=1}^n \sqrt{m_\pi^2+\bk_i^2},
\end{eqnarray}
where $\bp_1+\bp_2+\sum_{i=1}^n\bk_i = 0$.  The corresponding free hamiltonian is denoted by $H_0^n$.
Note that $E_W^n$ cannot be determined form the total energy $W$ alone, except for  the elastic scattering at $n=0$, where
$E_{W}^0$ is uniquely determined from a given value of $W$. 
Since the determination of $E_W^n$ from $W$ is important to construct potentials from the Schr\"odingier equation and $E_W^n$ for $n\ge 1$ cannot determined from $W$ in general, 
we restrict our considerations in this paper to cases where all momenta $\bp_1,\bp_2,\bk_1,\bk_2,\cdots,\bk_n$ are non-relativistic, so that we can write  $W \simeq W_{\rm th}^k + E^k_{W}$ for $k=1,2,\cdots, n$ at $W\in \Delta_n$. (We can exclude $k=0$ case since $E_W^0$ can always be determined from $W$ without non-relativistic approximation.)
This condition is explicitly written as $ \bp_i^2 < m_N^2$ for $i=1,2$ and $\bk_i^2 < m_\pi^2$ for $i=1,2,\cdots, n$.
Unless otherwise stated, we assume this condition in this paper.
We roughly estimate how many pions can be treated within this approximation. If the total energy of two nucleons with one pion at rest is equal to the minimum energy of $n$-pion production such that $2\sqrt{m_N^2+\bp^2} + m_\pi = 2 m_N + n m_\pi$, 
the non-relativistic condition, say  $\bp^2 \simeq 0.9\times  m_N^2$, leads to $ n-1 \le \frac{m_N}{m_\pi} (\sqrt{7.6}-2) \simeq  5$. Therefore we may consider up to $NN+6\pi$ with roughly 5\% relativistic corrections. 
Note that  some configurations of momenta may become relativistic for a given value of $W$.
We exclude such configurations in our consideration of this paper.

\subsection{Simplest case} 
\label{sec:NNpi}
To illustrate our strategy to construct energy-independent  potentials, let us consider the simplest case at $W < W_{\rm th}^2 = 2m_N + 2 m_\pi$ in this subsection. If $W \in \Delta_1$ ( $ 2m_N + m_\pi \le W <  2m_N + 2m_\pi$ ), the inelastic scattering with one pion production ($ NN\rightarrow NN+\pi$) becomes possible. We can define in this case a set of 4-independent equal time NBS wave function as
\begin{eqnarray}
Z_N \varphi_{W,c_0}^{00} (\bx_0) &=& \langle 0 \vert T\{ N(\bx,0) N(\bx+\bx_0,0) \} \vert NN, W, c_0\rangle_{\rm in}, \\
Z_N Z_\pi^{1/2} \varphi_{W,c_0}^{10} (\bx_0,\bx_1) &=& \langle 0 \vert  T\{N(\bx,0) N(\bx+\bx_0,0)\pi(\bx+\bx_1,0)\}\vert NN, W, c_0\rangle_{\rm in}, \\ 
Z_N \varphi_{W,c_1}^{01} (\bx_0) &=& \langle 0 \vert T\{N(\bx,0) N(\bx+\bx_0,0)\}\vert NN+\pi, W, c_1\rangle_{\rm in}, \\
Z_N Z_\pi^{1/2} \varphi_{W,c_1}^{11} (\bx_0,\bx_1) &=& \langle 0 \vert  T\{N(\bx,0) N(\bx+\bx_0,0)\pi(\bx+\bx_1,0) \}\vert NN+\pi, W, c_1\rangle_{\rm in} ,
\end{eqnarray}
where $Z_N$ and $Z_\pi$ are renormalization factors for nucleon and pion fields, such that
$ N(x) = Z_N^{1/2} N^r(x)$ and $\pi(x) = Z_\pi^{1/2} \pi^r(x)$, where $N^r(x)$ and $\pi^r(x)$ are renormalized nucleon and pion field, respectively. We here consider two asymptotic instates $\vert NN, W, c_0\rangle_{\rm in}$ and $\vert NN+\pi, W, c_1\rangle_{\rm in}$ corresponding to two nucleons  and two nucleons plus one pion, where $c_0$ and $c_1$ represent quantum numbers other than the total energy $W$. In the present case, $(W,c_0 )$  and $(W,c_1)$ are equivalent to $(s_1,s_2,\bp_1)$  and $(s_1,s_2,\bp_1,\bk_1)$ where $s_i$ is the helicity of the  $i$-th nucleon and $\bp_2$ is not independent due to the momentum conservation. As mentioned before, $W \simeq  W_0 + E_W^0 \simeq W_1 + E_W^1$.
If distances between all operators become large ($\vert\bx_0\vert, \vert \bx_1\vert, \vert \bx_1-\bx_0\vert \rightarrow \infty$),  we expect (and will indeed show in the separated paper\cite{HAL_NBS}) that all NBS wave functions given above satisfy free Schr\"odinger equations such that
\begin{eqnarray}
\left(E_{W}^0 -H_0^0\right)\varphi_{W,c_0}^{0i} &\simeq & 0 , \quad
\left(E_{W}^1 -H_0^1\right)\varphi_{W,c_0}^{1i} \simeq  0, \quad i=0,1 .
\end{eqnarray}

We consider the coupled channel Schr\"odinger equations for $NN$ and $NN+\pi$, which is given by
\begin{eqnarray}
(E_{W}^k - H_0^k) \varphi_{W,c_i}^{ki} &=& \sum_{l=0,1}\int \prod_{n=0}^l d^3 y_n\, U^{kl}([\bx]_k, [\by]_l) \varphi_{W,c_i}^{li}([\by]_l) , \quad k,i \in (0,1),
\label{eq:couple}
\end{eqnarray}
where $[\bx]_0 = \bx_0$ and $[\bx]_1=\bx_0,\bx_1$. 
Note that $E_W^1 \simeq W-W_{\rm th}^1 < 0 $ if $W\in \Delta_0$.
Our task is to show that $W$-independent $2\times 2$ potential matrix $U^{kl}$ exists.

For this purpose, we define vectors from these NBS wave functions at $W \in \Delta_1$  as
\begin{eqnarray}
\varphi_{W,c_i}^i &\equiv &
 \left( \varphi_{W,c_i}^{0i}([\bx]_0), \varphi_{W,c_i}^{1i}([\bx]_1)
\right)^T, \quad i=0,1 ,
\label{eq:vector1} 
\end{eqnarray}
while at $W\in \Delta_0$ we take only $\varphi_{W,c_0}^{0}$ as
\begin{eqnarray}
\varphi_{W,c_0}^0&\equiv &
 \left( \varphi_{W,c_0}^{00}([\bx]_0), \varphi_{W,c_0}^{10}([\bx]_1)
\right)^T, 
\label{eq:vector0}
\end{eqnarray}
where the second component $\varphi_{W,c_1}^{10}([\bx]_1)$ vanishes 
as distances between all operators go to infinity. (No asymptotic $NN+\pi$ state exists at $W < 2m_N+m_\pi$.)
Note that, instead of eq.~(\ref{eq:vector0}), we may define
\begin{eqnarray}
\varphi_{W,c_0}^0 &\equiv & 
 \left( \varphi_{W,c_0}^{00}([\bx]_0), 0 \right)^T, 
\label{eq:vector0a} 
\end{eqnarray}
at $W \in \Delta_0$.  Since  the definition of $\varphi_{W,c_0}^0$ at $W\in \Delta_0$ in eq.~(\ref{eq:vector0})  
will be required in Sec.~\ref{sec:t-dep} for  the time-dependent method,
we use it  in the main text of this paper, and  the construction with eq.~(\ref{eq:vector0a}) and other variations will be discussed in Appendix~\ref{app:comp}. 

As in the elastic case, we introduce the norm kernel  in the space spanned by $\varphi^i_{W,c_i}$ as
\begin{eqnarray}
{\cal N}^{ij}_{W_1c_i, W_2d_j} &=& \left( \varphi^i_{W_1,c_i}, \varphi^j_{W_2,d_j} \right) 
\equiv \sum_{k=0,1}\int \prod_{l=0}^k d^3x_l\,
 \overline{\varphi_{W_1,c_i}^{ki}([\bx]_k)} \varphi_{W_2,d_j}^{kj}( [\bx]_k) .
 \end{eqnarray}
 Here indices $i,j$ run over different ranges depending on values of $W_1,W_2$ such that
$i\in I(W_1)$ and $ j \in I(W_2)$, where $I(W) =\{ 0\}$ for $W\in \Delta_0$ and $I(W) =\{0,1\}$ for $W\in \Delta_1$.
Otherwise stated, we assume this in this subsection.

As long as $\varphi_{W,c_i}^i $ are  linearly independent, the Hermitian operator ${\cal N}$ has an inverse as
\begin{eqnarray}
\sum_{W\in \Delta_0+\Delta_1}\sum_{ h\in I(W),\, e_h} \, ( {\cal N}^{-1})_{W_1c_i,We_h}^{ih}\, {\cal N}_{We_h,W_2d_j}^{hj} 
&=& \delta^{ij}\delta_{W_1,W_2}\delta_{c_i,d_j} .
\end{eqnarray}
Schematically ${\cal N}$ has a following structure:
\begin{eqnarray}
{\cal N} &=& \left( 
\begin{array}{ccc}
{\cal N}^{00}(\Delta_0,\Delta_0),  & {\cal N}^{00}(\Delta_0,\Delta_1),  & {\cal N}^{01}(\Delta_0,\Delta_1) \\
 {\cal N}^{00}(\Delta_1,\Delta_0),  & {\cal N}^{00}(\Delta_1,\Delta_1),  &{\cal N}^{01}(\Delta_1,\Delta_1) \\
 {\cal N}^{10}(\Delta_1,\Delta_0),  &{\cal N}^{10}(\Delta_1,\Delta_1),  &{\cal N}^{11}(\Delta_1,\Delta_1)   \\
\end{array}
\right)
\end{eqnarray}
where ${\cal N}^{ab}(\Delta_i,\Delta_j)$ represent  a sub-matrix whose components are given by ${\cal N}^{ab}_{W_ic_a, W_jd_b}$ with $W_i\in \Delta_i$ and $W_j\in \Delta_j$ for $i,j,a,b=0$ or $1$. 
The corresponding inverse ${\cal N}^{-1}$ has of course the same structure.

Using this inverse, we define the ket vector $\vert \varphi_{W,c_i}^i\rangle$  and the corresponding bra vector $\langle  \psi_{W,c_i}^i \vert $, whose $k$-th components are given by
\begin{eqnarray}
\langle [\bx]_k \vert \varphi_{W,c_i}^i\rangle &=& \varphi^{ki}_{W,c_i}([\bx]_k), \\
\langle  \psi_{W,c_i}^i \vert [\bx]_k\rangle &=& \sum_{W_1\in \Delta_0\cup\Delta_1}\sum_{j\in I(W_1),d_j} ({\cal N}^{-1})^{ij}_{Wc_i,W_1d_j}
\overline{\varphi_{W_1,d_j}^{kj}([\bx]_k)}
\end{eqnarray}
for $k=0,1$, where $d_j$ runs over states which satisfies non-relativistic condition.
It is then easy to see that
\begin{eqnarray}
\langle \psi_{W_1,c_i}^i\vert  \varphi_{W_2,d_j}^j \rangle &=&
\sum_{k=0,1} \int \prod_{l=0}^k d^3x_l\, \langle \psi_{W_1,c_i}^{i} \vert [\bx]_k\rangle \langle [\bx]_k \vert
\varphi_{W_2,d_j}^{j}\rangle = ({\cal N}^{-1}\cdot {\cal N})^{ij}_{W_1c_i,W_2d_j} \nonumber \\
&=&\delta^{ij}\delta_{W_1,W_2}\delta_{c_i,d_j} .
\end{eqnarray}

Introducing operators $E_W$, $H_0$ and $U$ such that
\begin{eqnarray}
\langle [\bx]_k \vert (E_W - H_0)\vert [\by]_l\rangle &\equiv& 
\delta_{kl}(E_W^k -H_0^k)\prod_{n=0}^k\delta^{(3)}(\bx_n-\by_n)  
\label{eq:EH}\\
\langle [\bx]_k \vert U \vert [\by]_l \rangle &\equiv & U^{kl}([\bx]_k,[\by]_l), 
\label{eq:U}
\end{eqnarray}
the coupled channel Shcr\"odinger equation (\ref{eq:couple}) can be compactly written as\footnote{
Here and hereafter the sum over $c_i$ with $i\not=0$ is alway restricted to non-relativistic states if the number of particles is more than 2.}
\begin{eqnarray}
(E_W-H_0)\vert \varphi^i_{W,c_i}\rangle &=& U \vert \varphi^i_{W,c_i} \rangle .
\end{eqnarray}
Now it is easy to construct $U$ which satisfies the above equation as
\begin{eqnarray}
U &=& \sum_{W\in \Delta_0\cup\Delta_1}\sum_{i\in I(W)}\sum_{c_i} (E_W-H_0) \vert \varphi_{W,c_i}^i\rangle \langle \psi_{W,c_i}^i \vert ,
\label{eq:construction}
\end{eqnarray}
since
\begin{eqnarray}
U\vert \varphi_{W,c_i}^i\rangle &=& \sum_{W_1\in \Delta_0\cup\Delta_1}\sum_{j\in I(W_1)}\sum_{d_j} (E_W-H_0) \vert \varphi_{W_1,d_j}^j\rangle \langle \psi_{W_1,d_j}^j \vert  \varphi_{W,c_i}^i\rangle = (E_W-H_0) \vert \varphi_{W,c_i}^i\rangle .
\end{eqnarray}
An energy-independent potential matrix $U$ indeed exists. Note that $U$ is not unique since, for example, one can use
eq.~(\ref{eq:vector0a}) instead of eq.~(\ref{eq:vector0}) for $\varphi_{W,c_i}^i$, so that the resulting potential from eq.~(\ref{eq:construction}) differs from the one with eq.~(\ref{eq:vector0}).

Finally let us consider the Hermiticity of $U$. A matrix element of $U$  is evaluated as
\begin{eqnarray}
U^{ij}_{W_1c_i,W_2d_j} &\equiv & \langle \varphi_{W_1,c_i}^i\vert U \vert \varphi_{W_2,d_j}^j\rangle 
= \langle \varphi_{W_1,c_i}^i \vert (E_{W_2} - H_0) \vert \varphi_{W_2,d_j}^j\rangle , 
\end{eqnarray}
while 
\begin{eqnarray}
(U^\dagger)^{ij}_{W_1c_i,W_2d_j} &=&  
\overline{\langle \varphi_{W_2,d_j}^j \vert (E_{W_1} - H_0) \vert \varphi_{W_1,c_i}^i\rangle} 
=  \langle \varphi_{W_1,c_i}^i \vert (E_{W_1} - H_0) \vert \varphi_{W_2,d_j}^j\rangle .
\end{eqnarray}
Therefore potential $U$ is not Hermite in general. However it is effectively Hermite since in practice 
we solve the Schr\"odinger equation under the condition that $E_{W_1} = E_{W_2}$, which is equivalent to $W_1=W_2$ in our non-relativistic approximation. 

\subsection{General cases}
\label{sec:general}
It is not so difficult to extend the argument in the previous subsection to more general cases, where the total energy satisfies $W < W_{\rm}^{n_{\rm max}+1}$. As discussed before, the validity of the non-relativistic approximation requires $n_{\rm max}=5$
at most.

Let us consider  $W \in \Delta_0\cup \Delta_1\cup \cdots \cup \Delta_{n_{\rm max}}$.
At $W \in \Delta_s$ with $s\le n_{\rm max}$, we define a set of the equal time NBS wave functions as
\begin{eqnarray}
Z_N Z_\pi^{k/2}\varphi_{W,c_i}^{ki}([\bx]_k) &=& 
 \langle 0 \vert T\{N(\bx,0)N(\bx+\bx_0,0)\prod_{l=1}^k \pi(\bx+\bx_l,0)\}\vert NN+i\pi,W,c_i\rangle_{\rm in},  \  i \le s, 
 \nonumber \\
 &=& 0, \qquad i > s, 
 \nonumber \\
 \label{eq:NBS_general0}
\end{eqnarray}
where indices $k,i$ run from 0 to $n_{\rm max}$, but $\varphi_{W,c_i}^{ki}([\bx]_k)$ with $k > s$ vanishes, as distances among all operators (two nucleons and $k$ pions) becomes large,
$[\bx]_k = \bx_0,\bx_1,\cdots,\bx_k$ and $c_i$ represents quantum number other than the total energy $W$ of the instate.
In the present case, $(W,c_i)$ are equivalent  to $s_1,s_2,\bp_1,\bk_1,\bk_2,\cdots,\bk_i$ where $s_l$ is a helicity of the $l$-th nucleon. 

The coupled channel Schr\"odinger equation  for this system  at $W\in \Delta_s$ ($s\le n_{\rm max}$) is given by
\begin{eqnarray}
(E^k_W - H_0^k) \varphi_{W,c_i}^{ki} ([\bx]_k) &=& \sum_{l=0}^{n_{\rm max}} \int d [\by]_l \, U^{kl}([\bx]_k,[\by]_l)
\varphi_{W,c_i}^{li}([\by]_l), \quad  i \in I(W)
\label{eq:couple_general}
\end{eqnarray}
where $d[\by]_l =\displaystyle\prod_{m=0}^l d^3 y_m$, $I(W) = \{ 0,1,\cdots,s \}$ for $W\in \Delta_s$,
and $k =0,1,\cdots, n$. Note that $E_W^k \simeq W - W_{\rm th}^k < 0$ if $k\notin I(W)$.
It is now clear that the non-relativistic condition is necessary here to determine $E_W^k$ from $W,c_i$ if $k\not= i$.
Our task is to show that a $W$-independent $(n_{\rm max}+1)\times (n_{\rm max}+1)$ potential matrix $U$ exists.

As in the previous subsection, we define vectors of NBS wave functions with  $(n_{\rm max}+1)$ components at $W\in \Delta_s$ as
\begin{eqnarray}
\varphi^{i}_{W,c_i} &\equiv & \left(
 \varphi_{W,c_i}^{0i}([\bx]_0),  \varphi_{W,c_i}^{1i}([\bx]_1),\cdots,  \varphi_{W,c_i}^{n_{\rm max}i}([\bx]_{n_{\rm max}}) \right)^T,   
\end{eqnarray}
where $i$ runs over $I(W)$. 

The norm kernel is defined by
\begin{eqnarray}
{\cal N}_{W_1c_i,W_2d_j}^{ij} &\equiv& \left( \varphi_{W_1,c_i}^i,\varphi_{W_2,d_j}^j\right)
= \sum_{k=0}^{n_{\rm max}} \int d[\bx]_k \overline{\varphi_{W_1,c_i}^{ki}([\bx]_k)}\, \varphi_{W_2,d_j}^{kj}([\bx]_k),
\end{eqnarray}
whose inverse is denoted by ${\cal N}^{-1}$, where $i\in I(W_1)$ and $j\in I(W_2)$.
The bra and ket vectors, defined by
\begin{eqnarray}
\langle [\bx]_k\vert \varphi_{W,c_i}^i \rangle &=& \varphi_{W,c_i}^{ki}([\bx]_k), \\
\langle \psi_{W,c_i}^i\vert  [\bx]_k\rangle &=& \sum_{W_1}\sum_{j\in I(W_1)}\sum_{d_j}
({\cal N}^{-1})^{ij}_{Wc_i,W_1d_j}\overline{\varphi_{W_1,d_j}^{kj}([\bx]_k)}, 
\end{eqnarray}
satisfy
\begin{eqnarray}
\langle \psi_{W_1,c_i}^i\vert \varphi_{W_2,d_j}^j\rangle 
&=&\sum_{k=0}^{n_{\rm max}} \int d[\bx]_k \langle \psi_{W_1,c_i}^i\vert  [\bx]_k\rangle
\langle [\bx]_k\vert \varphi_{W_2,d_j}^j \rangle =\delta^{ij}\delta_{W_1,W_2}\delta_{c_i,d_j} .
\end{eqnarray}

Introducing operators $E_W$, $H_0$ and $U$, defined as in eqs.~(\ref{eq:EH}) and (\ref{eq:U}), 
we can construct
\begin{eqnarray}
U &=& \sum_{W}\sum_{i\in I(W)}\sum_{c_i}
(E_W-H_0) \vert \varphi_{W,c_i}^i\rangle \langle  \psi_{W,c_i}^i\vert ,
\end{eqnarray}
which satisfies
the coupled channel equation
\begin{eqnarray}
(E_W- H_0)\vert \varphi_{W,c_i}^i\rangle &=& U \vert \varphi_{W,c_i}^i\rangle .
\end{eqnarray}
It is also easy to see the effective Hermiticity of $U$: $U^{ij}_{W_1c_i,W_2d_j} =(U^\dagger)^{ij}_{W_1c_i,W_2d_j}$ at $W_1=W_2$ (with non-relativistic approximation).

\subsection{Special case without non-relativistic approximation}
\label{sec:special}
In this subsection, we discuss a special case of inelastic scatterings where non-relativistic approximation is not required to construct energy independent potentials. Here, coupled two-particle scattering channels such as $A_i B_i\rightarrow A_jB_j $ with $i,j = 1,2,\cdots, n_{\rm max}$ are considered. 
For example, in the baryon scattering in the strangeness $S=-2$ and isospin $I=0$ channel, $\Lambda\Lambda, N\Xi$ and $\Sigma\Sigma$ appear as asymptotic states if the total energy $W$ in the center of mass system is 
$2 m_\Sigma \le W < 2m_\Sigma + m_\pi$.
The method to extract coupled channel potentials in this kind of situation has already been proposed in Ref.~\cite{Aoki:2011gt}, under an assumption that energy independent coupled channel potentials exist.
In this subsection we prove this assumption.

Given the total energy $W$, the relative momentum $\bp_i$(squared) and the kinetic energy $E_W^i$, together with the free Hamiltonian $H_0$,  for $A_iB_i$ are given by 
\begin{eqnarray}
W = \sqrt{\bp_i^2 + m_{A_i}^2} + \sqrt{\bp_i^2 + m_{B_i}^2}, \qquad
E_W^i = \frac{\bp_i^2}{2 m_r^i}, \ H_0 = \frac{-\nabla^2}{2 m_r^i},  \quad m_r^i = \frac{m_{A_i} m_{B_i}}{m_{A_i}+m_{B_i}},
\end{eqnarray}
where $m_{A_i}$ and $m_{B_i}$ are masses of $A_i$ and $B_i$, and $m_r^i$ is their reduced mass.
We here assume $m_{A_i} + m_{B_i} < m_{A_j} + m_{B_j}$ for $i < j$.  
Note that if $W < W_{\rm th}^i\equiv m_{A_i}+m_{B_i}$, $\bp_i^2$ and $E_W^i$ become negative.

We defined NBS wave function for $A_k B_k$  as
\begin{eqnarray}
(Z_{A_k} Z_{B_k})^{1/2} \varphi_{W,c_i}^{ki} (\bx) &=& \langle 0 \vert  T\{A_k(\br,0) B_k(\br+\bx,0)\} \vert A_iB_i, W, c_i\rangle_{\rm in}, 
\end{eqnarray}
where $Z_{A_k},Z_{B_k}$ are renormalization factors defined by $ A_k(x) = Z_{A_k}^{1/2} A_k^r(x)$ and $ B_k(x) = Z_{B_k}^{1/2} B_k^r(x)$ with bare fields $A_k,B_k$ and renormalized fields $A_k^r,B_k^r$, and $c_i$ represents quantum number of the asymptotic instate $\vert A_iB_i,W,c_i\rangle_{\rm in}$ other than $W$.
The index $k$ always runs from 1 to $n_{\rm max}$, while the index $i$ runs over $I(W) = 1,2,\cdots, s-1$ if $W_{\rm th}^{s-1} \le W < W_{\rm th}^s$. 
We can show that 
\begin{eqnarray}
\lim_{\vert \bx\vert\rightarrow \infty} (E_W^k - H_0)  \varphi_{W,c_i}^{ki} (\bx) = 0 ,
\end{eqnarray}
and $\varphi_{W,c_i}^{ki} (\bx) $ carries the information of scattering phase shifts~\cite{Aoki:2011gt}.

We define vectors $ \vert \varphi_{W,c_i}^i\rangle$ and the corresponding norm kernel as
\begin{eqnarray}
\langle \bx, k \vert \varphi_{W,c_i}^i\rangle &=& \varphi_{W,c_i}^{ki}(\bx), \\
{\cal N}^{ij}_{W_1c_i,W_2c_j} &= & (\varphi_{W_1,c_i}^i, \varphi_{W_2,c_j}^j) \equiv
\sum_{k=1}^{n_{\rm max}} \int  d^3x\, \overline{\varphi_{W_1,c_i}^{ki}(\bx)} \ \varphi_{W_2,c_j}^{kj} (\bx),
\end{eqnarray}
where $i\in I(W_1)$ and $j\in I(W_2)$.
Using the inverse ${\cal N}^{-1}$ of ${\cal N}$, we construct dual vectors
\begin{equation}
\langle \psi_{W, c_i}^i\vert \bx,k \rangle =\sum_{W_1,j\in I(W_1),c_j} \left( {\cal N}^{-1}\right)^{ij}_{Wc_i,W_1c_j}\overline{\langle \bx,k \vert \varphi_{W_1,c_j}^j\rangle} ,
\end{equation}
which satisfies
\begin{eqnarray}
\sum_{k=1}^{n_{\rm max}} \int d^3x\, \langle \psi_{W_1, c_i}^i\vert \bx,k \rangle \cdot  \langle \bx,k \vert  \varphi_{W_2, d_j}^j \rangle
&=& \delta^{ij}\delta_{W_1,W_2}\delta_{c_i,d_j} .
\end{eqnarray}

An energy independent $n_{\rm max}\times n_{\rm max}$ potential matrix which satisfies the coupled channel equation that
\begin{eqnarray}
(E^k_W - H_0)\varphi_{W,c_i}^{ki}(\bx) &=& \sum_{l=1}^{n_{\rm max}} \int d^3y\, U^{kl}(\bx,\by) \varphi_{W,c_i}^{li}(\by),
\end{eqnarray}
can be constructed as
\begin{eqnarray}
U^{kl}(\bx,\by) &=& \sum_{W, i\in I(W), c_i} (E_W^k - H_0)\langle \bx,k\vert \varphi_{W,c_i}^i\rangle \langle
\psi_{W,c_i}^i \vert \by, l\rangle ,
\end{eqnarray}
which is manifestly energy ($W$) independent, and is Hermite at fixed $W$.
  
\section{Time dependent method}
\label{sec:t-dep}
In Ref.~\cite{HALQCD:2012aa}, a method to extract hadronic potentials below inelastic thresholds from
time dependence of correlation functions has been proposed, in order to overcome difficulties in the conventional method where NBS wave functions with definite energies are extracted from asymptotic behaviors of correlation functions in time.   In this section, we extend the method so that it can be applicable to the case above inelastic thresholds.

The normalized correlation function  is defined by
\begin{eqnarray}
Z_N Z_\pi^{k/2} R^k([\bx]_k,t) &=& \frac{1}{e^{-W_{\rm th}^k t}} 
\langle 0 \vert T\{N(\bx,t) N(\bx+\bx_0,t)\prod_{l=1}^k\pi(\bx+\bx_l,t) {\cal J}_{NN}
(0)\}\vert 0\rangle
\end{eqnarray}
for $k=0,1,2,\cdots, n_{\rm max}$,
where ${\cal J}_{NN}$ is some source operator which couples to $NN$ states.  Inserting the complete set for the $NN$ system that
\begin{eqnarray}
{\bf 1} &=& \sum_{W} \sum_{i\in I(W)} \sum_{c_i} \vert NN + i\pi, W,c_i \rangle_{\rm in}\ {}_{\rm in}\langle
NN+i\pi,W,c_i\vert +\cdots,
\end{eqnarray}
where the ellipsis represents states with $W > W_{\rm th}^{n_{\rm max}}$ and are neglected hereafter,
into the above correlation function, we obtain
\begin{eqnarray}
R^k([\bx]_k,t) &= & \sum_{W} \sum_{i\in I(W)}\sum_{c_i}e^{-\Delta^k W\, t}\varphi_{W,c_i}^{ki}([\bx]_k)
A_{W,c_i}^i,
\end{eqnarray}
where
\begin{eqnarray}
A_{W,c_i}^i &=& {}_{\rm in}\langle NN+i\pi,W,c_i\vert {\cal J}_{NN}(0)\vert 0\rangle, \quad
\Delta^k W \equiv W - W_{\rm th}^k \simeq E_W^k .
\end{eqnarray}
Note that $R^k$ automatically contains a sum over $W,i\in I(W), c_i$, which is necessary to define the non-local potentials in the previous section but is difficult in practice to perform one by one.
Note however that states with relativistic momenta may appear in the sum.
We here assume that contributions from such states can be suppressed by an appropriate choice of ${\cal J}_{NN}$. 
Using the non-relativistic approximation, we can derive
\begin{eqnarray}
\left\{-H_0^k -\frac{\partial}{\partial t}\right\}\cdot R^k([\bx]_k,t) &=&\sum_{W,i\in I(W),c_i} e^{-\Delta^k W t}
\sum_{l=0}^{n_{\rm max}}  \int d[\by]_l\, U^{kl}([\bx]_k,[\by]_l) \varphi_{W,c_i}^{li}([\by]_l)A_{W,c_i}^i \nonumber \\
&=& \sum_{l=0}^{n_{\rm max}}  e^{-(l-k) m_\pi t} \int d[\by]_l U^{kl}([\bx]_k,[\by]_l) \sum_{W,i\in I(W),c_i}e^{-\Delta^l W t}\varphi^{li}([\by]_l)
A_{W,c_i}^i \nonumber \\
&=& \sum_{l=0}^{n_{\rm max}}  e^{-(l-k) m_\pi t} \int d[\by]_l U^{kl}([\bx]_k,[\by]_l) R^l([\by]_l,t) .
\end{eqnarray}
We then finally obtain
\begin{eqnarray}
\left\{-H_0^k -\frac{\partial}{\partial t}\right\}\cdot R^k([\bx]_k,t) &=& 
e^{k m_\pi t}\sum_{l=0}^{n_{\rm max}}  e^{- l m_\pi t} \int d[\by]_l U^{kl}([\bx]_k,[\by]_l) R^l([\by]_l,t) ,
\end{eqnarray}
which can be used to obtain $U^{kl}$, combined with the derivative expansion~\cite{HALQCD:2012aa}.

We here propose a method to extract $U^{kl}$ directly. For this purpose, we consider a set of more complicated correlation functions defined by
\begin{eqnarray}
R^{kl}([\bx]_k,[\by]_l,t) &=& \frac{1}{e^{-W_{\rm th}^k t}}
\langle 0 \vert  T\{N(\bx,t)N(\bx+\bx_0,t)\prod_{m=1}^k \pi(\bx+\bx_m,t)\nonumber \\
&\times& \int d^3y\, \bar N(\by,0) \bar N(\by+\by_0,0)\prod_{s=1}^l \pi^\dagger(\by+\by_s,0)\}\vert 0\rangle,
\end{eqnarray}
which satisfies
\begin{eqnarray}
\left\{-H_0^k -\frac{\partial}{\partial t}\right\}\cdot R^{kl}([\bx]_k,[\by]_l,t) &=&
\sum_{s=0}^{n_{\rm max}}  e^{-(s-k)m_\pi t} \int d[\bz]_s U^{ks}([\bx]_k,[\bz]_s) R^{sl}([\bz]_s,[\by]_l,t). 
\end{eqnarray}
Using real eigenvalues $\lambda_m$ of  the Hermitian operator $R$ and their eigenvectors $v_m$ whose $k$-th component is given by $v_m^k([\bx]_k,t)$ with $m=0,1,\cdots, {n_{\rm max}} $, we can construct the inverse of $R$ as 
\begin{eqnarray}
(R^{-1})^{kl}([\bx]_k,[\by]_l,t) &=& \sum_{m=n_0}^{n_{\rm max}}  \frac{1}{\lambda_m} v_m^k([\bx]_k,t)\, \{ v_m^l([\by]_l,t)\}^\dagger .
\end{eqnarray}
Note that we remove zero modes with $\lambda_m =0$ from $R$ and $R^{-1}$, so that the dimension of $R$ and $R^{-1}$ are effectively reduced from $({n_{\rm max}} +1) \times ({n_{\rm max}} +1)$ to $({n_{\rm max}} +1-n_0)\times ({n_{\rm max}} +1-n_0)$ where $n_0$ is the number of zero modes. 

Using the inverse $R^{-1}$, we can extract $U$ as
\begin{eqnarray}
U^{kl}([\bx]_k,[\by]_l) &=& e^{-k m_\pi t} \sum_{s=0}^{n_{\rm max}} \int d[\bz]_s  \left\{-H_0^k -\frac{\partial}{\partial t}\right\}\cdot R^{ks}([\bx]_k,[\bz]_s,t) (R^{-1})^{sl}([\bz]_s,[\by]_l,t)e^{lm_\pi t} . \nonumber \\
\end{eqnarray}

\section{Conclusion and discussion}
\label{sec:conclusion}
In this paper,  we have shown that energy independent and non-local potentials can be constructed from a particular set of NBS wave functions even above inelastic thresholds as long as momenta of all particles involved are non-relativistic (Sec.~\ref{sec:NNpi} and Sec.~\ref{sec:general}) or  a number of particles  is always two (Sec.~\ref{sec:special}). We have also derived a formula to extract non-local potentials with non-relativistic approximations using the time dependent method proposed in Ref.~\cite{HALQCD:2012aa}.

By the same method in Sec.~\ref{sec:NNpi} and Sec.~\ref{sec:general}, we can construct  
an energy independent non-local potential for three-nucleon systems\cite{Doi:2010yh,Doi:2011gq} and even for systems with more than 3 nucleons.
In the case of inelastic scattering such as $\Lambda\Lambda \rightarrow \Lambda\Lambda, N\Xi, \Sigma\Sigma$
\cite{Sasaki:2010bh}, the result in Sec.~\ref{sec:special} has completed the HALQCD method proposed in Ref.~\cite{Aoki:2011gt}, where non-relativistic approximation is not required.

An existence of energy independent potentials, which is one of the important properties necessary for the HALQCD method to investigate hadronic interactions, is now established in rather general situations.
A remaining important property to be proven is an asymptotic behavior of NBS wave functions for more than 2 particles and its relation to $S$-matrix of the corresponding quantum field theory.
Results on this issue will be published elsewhere\cite{HAL_NBS}.

\section*{Acknowledgement}

S.A. would like to thank the Institute for Nuclear Theory at the University of Washington for its hospitality and the Department of Energy for partial support during the completion of this work. He also thanks Profs. M. Savage and S. Beane for fruitful discussions, which initiated this investigation.
This work is supported in part by the Grants-in-Aid for Scientific Research (No. 24740146), the Grant-in-Aid for Scientific Research on Innovative Areas (No. 2004: 20105001, 20105003) and SPIRE (Strategic Program for Innovative Research).

\appendix
\section{Comparisons among different constructions}
\label{app:comp}
The energy-independent $(n_{\rm max} +1)\times (n_{\rm max}+1)$ potential matrix  in the main text is given in the coordinate space by
\begin{eqnarray}
U^{kl}&=& \sum_W\sum_{i\in I(W)}\sum_{c_i} (E_W^k - H_0^k)\vert \varphi_{W,c_i}^{ki}\rangle \langle \psi_{W,c_i}^{li}\vert 
\label{eq:original}
\end{eqnarray}
for $0\le k,l \le n_{\rm max}$, where $\langle [\bx]_k\vert \varphi^{ki}_{W,c_i} \rangle= \varphi_{W,c_i}^{ki}([\bx]_k)$ and 
 $\sum_{k=0}^{n_{\rm max}} \langle \psi^{ki}_{W_1,c_i}\vert \varphi^{kj}_{W_2,d_j}\rangle =\delta^{ij}\delta_{W_1,W_2}\delta_{c_i,d_j}$.
The corresponding coupled channel Schr\"odingier equation is  given by
\begin{eqnarray}
(E_W - H_0) \vert \varphi_{W,c_i}^{ki}\rangle &=&
\sum_{l=0}^{n_{\rm max}}  U^{kl}\vert \varphi_{W,c_i}^{li}\rangle
\label{eq:original_eq}
\end{eqnarray}
for $0\le k \le n_{\rm max}$ and $i \in I(W)$.

In this appendix we consider some other constructions of energy-independent potentials in terms of NBS wave functions and compare them with eq.~(\ref{eq:original}).

\subsection{Modified wave function vectors}
\label{app:modified}
As already mentioned in the main text, we can define the vectors of NBS wave functions using eq.~(\ref{eq:vector0a})
instead of eq.~(\ref{eq:vector0}). The corresponding modification to eq.~(\ref{eq:NBS_general0}) becomes
\begin{eqnarray}
Z_N Z_\pi^{k/2}\varphi_{W,c_i}^{ki}([\bx]_k) &=& 
 \langle 0 \vert T\{N(\bx,0)N(\bx+\bx_0,0)\prod_{l=1}^k \pi(\bx+\bx_l,0)\}\vert NN+i\pi,W,c_i\rangle_{\rm in},  \  k,i \le s, 
 \nonumber \\
 &=& 0, \qquad {\rm otherwise}.
 \nonumber \\
 \label{eq:NBS_general0a}
\end{eqnarray}
The energy-independent potential $U^{kl}_M$ (where $M$ represents "modified")  is given by the same formula in eq.~(\ref{eq:original}) with modifications by  (\ref{eq:NBS_general0a}) to $\vert\varphi_{W,c_i}^i\rangle$ and $\langle \psi_{W,c_i}^i\vert$, while the corresponding Schr\"odingier equation reads
\begin{eqnarray}
(E_W - H_0) \vert \varphi_{W,c_i}^{ki}\rangle &=&
\sum_{l\in I(W)}  U_M^{kl}\vert \varphi_{W,c_i}^{li}\rangle
\label{eq:modified_eq}
\end{eqnarray}
 for $k,i \in I(W)$, where $I(W)=0,1,2,\cdots, s$ at $W\in \Delta_s =[W^s_{\rm th}, W^{s+1}_{\rm th})$.

\subsection{Recursive construction}
\label{app:recursion}
We construct another energy-independent potential recursively starting from the potential for the elastics $NN$ scattering. 

We first define the $U_R^{00}$ corresponding to $NN\rightarrow NN$ elastic scattering as
\begin{eqnarray}
U_R^{00} &=& \sum_{W\in \Delta_0}\sum_{c_0} (E_W-H_0)  \vert \varphi_{W,c_0}^{00}\rangle 
 \langle \psi_{W,c_0}^{00}\vert 
\label{eq:recursive_U00}
\end{eqnarray}
where the dual wave function $\langle \psi_{W,c_0}^{00}\vert $ to  $\vert \varphi_{W,c_0}^{00}\rangle$ 
satisfies $\langle \psi_{W_1,c_0}^{00}\vert \varphi_{W_2,d_0}^{00}\rangle =\delta_{W_1,W_2}\delta_{c_0,d_0}$ at $W_1,W_2\in \Delta_0$.
This $U_R^{00}$ is identical to the elastic potential given  in eq.~(\ref{eq:potentialW}) and satisfies
\begin{eqnarray}
(E_W - H_0) \vert \varphi_{W,c_0}^{00} \rangle &=&
 U^{00}_R \vert  \varphi_{W,c_0}^{00}\rangle
\label{eq:recursive_eq00}
\end{eqnarray}
at $ W\in \Delta_0$.

We then increase the energy so that $W\in \Delta_1$. A condition that $\varphi^{ki}_{W,c_i}$ for $ 0\le k,i \le 1$ satisfy the corresponding Schr\"dingier equation leads to
 \begin{eqnarray}
U_R^{01} &=& \sum_{W\in \Delta_1}\sum_{i=0,1}\sum_{c_i}\left[ (E_W-H_0) \vert \varphi_{W,c_i}^{0i} \rangle- U^{00}_R\vert \varphi^{0i}_{W,c_i}\rangle \right] 
 \langle \psi_{W,c_i}^{1i}\vert,
\label{eq:recursive_U01}
\end{eqnarray}
where $\langle \psi_{W,c_i}^{1i}\vert$ for $i=0,1$ satisfy $\langle\psi_{W_1,c_i}^{1i}\vert \varphi_{W_2,d_j}^{1j}\rangle=\delta^{ij}\delta_{W_1,W_2}\delta_{c_i,d_j}$ at $W_1,W_2\in \Delta_1$.
Note that $U_R^{00}$ used here is determined in eq.~(\ref{eq:recursive_eq00}) at  the elastic region.
We define $U_R^{10}$ by imposing hermiticity for the potential, i.e. $U_R^{10} =( U_R^{01})^\dagger$, from which
we can finally determine  
\begin{eqnarray}
U_R^{11}&=& \sum_{W\in \Delta_1}\sum_{i=0,1}\sum_{c_i}\left[ (E_W-H_0)  \vert\varphi_{W,c_i}^{1i} 
\rangle - U^{10}_R\vert \varphi^{0i}_{W,c_i}\rangle \right] \langle \psi_{W,c_i}^{1i}\vert.
\label{eq:recursive_U11}
\end{eqnarray}
We now have $U^{00}_R$ at $W\in \Delta_0$ and $U_R^{ij}$ for $0\le i,j \le 1$ at $W\in \Delta_1$.

It is not so difficult to extend the above construction to larger $W$ recessively.
We assume that the $s\times s$ potential matrix $U_R^{ij}$ is already determined at $W\in \Delta_{s-1}$
for $s\le n_{\rm max}$. At $W\in \Delta_s$, $U_{ks}$ for $k<  s$ can be obtained by
\begin{eqnarray}
U_R^{ks}  &=& \sum_{W\in \Delta_s}\sum_{i\in I(W)}\sum_{c_i}\left[ (E_W-H_0)\vert \varphi_{W,c_i}^{ki}\rangle -\sum_{l=0}^{s-1}U_R^{kl}\vert \varphi_{W,c_i}^{li}\rangle\right]
\langle \psi_{W,c_i}^{si}\vert,
\label{eq:recursive_Uks}
\end{eqnarray}
where $\langle \psi^{sj}_{W,c_i}\vert $ for $i=0,1,\cdots, s$ satisfy $\langle\psi^{si}_{W_1,c_i}\vert\varphi_{W_2,d_j}^{sj} \rangle =\delta^{ij}\delta_{W_1,W_2}\delta_{c_i,d_j}$ at $W_1,W_2\in \Delta_s$. Using the Hermiticity relation that $U_R^{sk}
= (U_R^{ks})^\dagger$ for $k=0,1,\cdots, s-1$,  we obtain $U^{ss}_R$ as
\begin{eqnarray}
U_R^{ss} &=& \sum_{W\in \Delta_s}\sum_{i\in I(W)}\sum_{c_i}\left[ (E_W-H_0)\vert \varphi_{W,c_i}^{si}\rangle -\sum_{l=0}^{s-1} U_R^{sl}\vert \varphi_{W,c_i}^{li}\rangle
\right] \langle\psi_{W,c_i}^{si}\vert .
\label{eq:recursive_Uss}
\end{eqnarray}
The $(s+1)\times (s+1)$ potential matrix $U^{kl}_R$ for $0\le k,l \le s$  is constructed. 
We can continue this recursive construction until $s=n_{\rm max}$.

The corresponding Schr\"odingier equation at $W\in \Delta_s$ becomes
\begin{eqnarray}
(E_W -H_0) \vert \varphi_{W,c_i}^{ki} \rangle &=&
\sum_{l=0}^s  U_R^{kl}\vert \varphi_{W,c_i}^{li}\rangle
\label{eq:recursive_eq}
\end{eqnarray}
for  $0\le k,i \le s$, where off-diagonal elements $U^{kl}$ for $k\not= l$ are Hermite by construction.

\subsection{Construction at each energy interval}
\label{app:interval}
We  finally give a construction of the potential matrix different at each energy interval.
At $W\in \Delta_s$ for $0\le s \le n_{\rm max}$,  the $(s+1)\times (s+1)$ potential matrix can be constructed as
\begin{eqnarray}
U_s^{kl}&=& \sum_{W\in \Delta_s}\sum_{i\in I(W)}\sum_{c_i} (E_W-H_0)\vert\varphi^{ki}_{W,c_i}\rangle\langle\psi_{W,c_i}^{li}\vert
\label{eq:indep_U}
\end{eqnarray}
for $ 0\le k,l \le s$, where $\langle \psi^{ki}_{W,c_i}\vert$ for $i=0,1,\cdots, s$ satisfy $\sum_{k=0}^s \langle\psi^{ki}_{W_1,c_i}\vert \varphi^{kj}_{W_2,d_j}\rangle =\delta^{ij}\delta_{W_1,W_2}\delta_{c_i,d_j}$ at $W_1,W_2\in \Delta_s$.
Note that $U_0^{00}$ is identical to $U^{00}$ given in eq.~(\ref{eq:potentialW}).

The corresponding Schr\"dingier equation at $W\in \Delta_s$ becomes
\begin{eqnarray}
(E_W - H_0) \vert\varphi_{W,c_i}^{ki}\rangle&=&
\sum_{l=0}^s  U_s^{kl}\vert\varphi_{W,c_i}^{li}\rangle
\label{eq:indep_eq}
\end{eqnarray}
for $0\le k, i \le s$. 

\subsection{Comparison}
\noindent
Properties of the original construction in the main text are as follows.
\begin{enumerate}
\item A size of the potential matrix $U^{kl}$ is always $(n_{\rm max}+1)^2$ at  all $W \in[ W_{\rm th}^0, W_{\rm th}^{n_{\rm max}+1})$.
\item A form of the potential matrix given in eq.~(\ref{eq:original}) is also same at all energy.
\item We use $(n_{\rm max} + 1)$-length vectors $\{ \vert \varphi^{0i}_{W,c_i}\rangle,
\vert \varphi^{1i}_{W,c_i}\rangle,\cdots, \vert \varphi^{n_{\rm max}i}_{W,c_i}\rangle \}$,
which are taken to be linearly independent  
for different values of $W \in[ W_{\rm th}^0, W_{\rm th}^{n_{\rm max}+1})$, $i\in I(W)$ and $ c_i$.
\item The construction can be combined with the time dependent method in Sec.~\ref{sec:t-dep}.
\end{enumerate}
In the case of the modified wave function vectors, we have
\begin{enumerate}
\item A size of the potential matrix $U^{kl}_M$ is $(s+1)^2$ at $W\in \Delta_s$.
\item The form of $U^{kl}_M$ is same at all energy where $U_M^{kl}$ is defined.  
\item We use vectors  $\{ \vert \varphi^{0i}_{W,c_i}\rangle,
\vert \varphi^{1i}_{W,c_i}\rangle,\cdots, \vert \varphi^{si}_{W,c_i}\rangle,0,\cdots, 0 \}$,
which are linearly independent
for different values of $W \in[ W_{\rm th}^0, W_{\rm th}^{n_{\rm max}+1})$, $i\in I(W)$ and $ c_i$. The (effective) length of these vectors is $s+1$ at $W\in \Delta_s$.
\end{enumerate}
For the recursive construction, we have
\begin{enumerate}
\item A size of the potential matrix $U^{kl}_R$ is $(s+1)^2$ at $W\in \Delta_s$.
\item The form of $U^{kl}_R$ is same at all energy where $U_R^{kl}$ is defined.  
\item We use  $\vert \varphi^{si}_{W,c_i}\rangle $, which are inearly independent for different values of $W \in \Delta_s$, $i\in I(W)$ and $ c_i$.
\item The potential matrix is recursively constructed:
At $W\in \Delta_s$, $U_R^{ks}$ for $k=0,1,2,\cdots, s$ are determined from $\{ U^{k^\prime s}\, | \ k^\prime < k\}$, while $U^{sk}_R$ can be obtained from $U_R^{ks}$ using Hermiticity.  
\end{enumerate}
For the construction in Appendix~\ref{app:interval}, we have
\begin{enumerate}
\item A size of the potential matrix $U^{kl}$ is $(s+1)^2$ at $W\in \Delta_s$.
\item The form of $U^{kl}_s$  is different for each $s$ at $W\in \Delta_s$.  
\item We use the $(s+1)$ length vectors $\{\vert \varphi^{0i}_{W,c_i}\rangle,\vert\varphi^{1i}_{W,c_i}\rangle,\cdots, \vert \varphi^{si}_{W,c_i}\rangle \}$, which are  linearly independent for different values of $W \in \Delta_s$, $i\in I(W)$ and $ c_i$.
\item $U_s^{kl}$ can be determined at each energy interval $\Delta_s$, without using information of other energy intervals.
\end{enumerate}
We summarize the above properties in table~\ref{tab:comparison}.
\begin{table}[htbdp]
\caption{A comparison of different constructions}
\begin{center}
\begin{tabular}{|c|c|c|c|c|}
\hline
construction & original & modified(App.~\ref{app:modified}) & recursive(App.~\ref{app:recursion}) & interval(App.~\ref{app:interval}) \\
\hline
size of $U$ at $W\in\Delta_s$ & $(n_{\rm max}+1)^2$ & $(s+1)^2$ & $(s+1)^2$ & $(s+1)^2$ \\
$\Delta_s$ dependence of $U$ & no & no & no & yes \\
vectors  &$\{ \vert \varphi^{k,i}_{W,c_i}\rangle \}_{k\le n_{\rm max}}$ & $\{\vert \varphi^{k, i}_{W,c_i}\rangle\}_{k\le s}$& $\vert \varphi^{s,i}_{W,c_i}\rangle$  &  $\{\vert \varphi^{k,i}_{W,c_i}\rangle \}_{k\le s}$ \\
feature & $t$-dependent method & & recursive & each $\Delta_s$\\
\hline
\end{tabular}
\end{center}
\label{tab:comparison}
\end{table}%

\end{document}